\documentclass[10pt, onecolumn]{extarticle}

\usepackage[bordercolor=white,backgroundcolor=gray!30,linecolor=black,colorinlistoftodos, textsize=10pt]{todonotes}

\usepackage{amssymb}
\usepackage{mathtools}
\usepackage{textcomp}
\usepackage{upgreek}
\usepackage{amsmath}
\usepackage[textstyle,amssymb]{SIunits}
\usepackage{graphicx}
\usepackage[normalem]{ulem}  
\usepackage{soul}
\usepackage[twocolumn,textwidth=183mm,columnsep=6mm,top=20mm,bottom=25mm]{geometry}
\usepackage{helvet}
\usepackage{titlesec}
\titleformat*{\section}{\bfseries\sffamily}
\titlespacing{\section}{0pt}{*4}{*0}
\titleformat{\subsection}[runin]{\normalfont\bfseries}{\thesubsection.}{3pt}{}
\usepackage{setspace}
\usepackage[breaklinks=true]{hyperref} 
\usepackage[hang,flushmargin]{footmisc} 
\usepackage[font={footnotesize,sf},labelfont={sf,bf},labelsep=endash,justification=raggedright]{caption}

\usepackage{natbib}
\makeatletter
\def\blfootnote{\gdef\@thefnmark{}\@footnotetext}
\makeatother

\usepackage[labelfont=bf,justification=justified]{caption}

\begin{document}

\twocolumn[
\begin{@twocolumnfalse}
	{\LARGE \sf \textbf{Battery-operated mid-infrared diode laser frequency combs}}
	\vspace{0.4cm}
	
	{\sf\large \textbf {Lukasz A. Sterczewski$^{1,2,*}$, Mathieu~Fradet$^{1}$, Clifford~Frez$^{1}$, Siamak~Forouhar$^{1}$, \\and Mahmood~Bagheri$^{1,\mathparagraph}$}}
		\vspace{0.5cm}

		{\sf \textbf {\noindent $^1$Jet Propulsion Laboratory, California Institute of Technology, Pasadena, CA 91109, USA}\\
		{\sf \textbf {\noindent $^2$Faculty of Electronics, Photonics and Microsystems, Wroclaw University of Science and Technology, Wyb. Wyspianskiego 27, 50-370 Wroclaw, Poland}
}}
		\vspace{0.5cm}
		
\vspace{0.1cm}
\footnotesize
This is the peer reviewed version of the following article: L. A. Sterczewski, M. Fradet, C. Frez, S. Forouhar, and M. Bagheri, "Battery-operated mid-infrared diode laser frequency combs," Laser \& Photonics Reviews, 2200224, (2023), which has been published in final form at \url{https://doi.org/10.1002/lpor.202200224}. This article may be used for non-commercial purposes in accordance with Wiley Terms and Conditions for Use of Self-Archived Versions. This article may not be enhanced, enriched or otherwise transformed into a derivative work, without express permission from Wiley or by statutory rights under applicable legislation. Copyright notices must not be removed, obscured or modified. The article must be linked to Wiley’s version of record on Wiley Online Library and any embedding, framing or otherwise making available the article or pages thereof by third parties from platforms, services and websites other than Wiley Online Library must be prohibited.
\vspace{0.5cm}
\normalsize
\end{@twocolumnfalse}]
\vspace{0.5cm}


{\noindent \sf \small \textbf{\boldmath
Mid-wave infrared (MIR, 3--5~\textmu m) optical frequency combs (OFC) are of critical importance for spectroscopy of fundamental molecular transitions in space and terrestrial applications. Although in this band OFCs can be obtained via supercontinuum or difference frequency generation using optical pumping schemes, unprecedented source miniaturization and monolithic design are unique to electrically-pumped semiconductor laser structures. To date, high-brightness OFC generation in this region has been demonstrated in quantum- and interband cascade lasers (QCL/ICL), yet with sub-optimal spectral properties. Here, we show a MIR quantum well diode laser (QWDL) OFC, whose excellent spectral uniformity, narrow optical linewidths, and milliwatt optical power are obtained at a fraction of a watt of power consumption. The continuously tunable source offers $\sim$1~THz of optical span centered at 3.04~$\upmu$m, and a repetition rate of 10~GHz. As a proof-of-principle, a directly-battery-operated MIR dual-comb source is shown with almost 0.5~THz of optical coverage accessible in the electrical domain in microseconds. These results indicate the high suitability of QWDL OFCs for future chip-based real-time sensing systems in the mid-infrared.}}

\blfootnote{\noindent$^*${lukasz.sterczewski@pwr.edu.pl}}
\blfootnote{\noindent$^\mathparagraph${mahmood.bagheri@jpl.nasa.gov}}

\section*{Introduction}
\noindent Access to the MIR spectral region (3--5~\textmu m) plays a vital role for molecular spectroscopy due to the presence of strong ro-vibrational transitions of many environmentally-important species in this band. High sensitivity and selectivity makes trace gas monitoring, breath analysis, and industrial quality control particularly attractive in the MIR. While spectrally-broadband emission in this region can be obtained using thermal incoherent sources such as globars~\cite{davis2001fourier}, revolutionary new coherent light emitters offer unmatched optical power and the discrete structure of an optical frequency comb (OFC)~\cite{hansch2006nobel}. The latter feature enables performing precise, broadband and high-resolution measurements in the moving-parts-free configuration of a dual-comb spectrometer (DCS)~\cite{coddington_dual-comb_2016}, which naturally links the optical and radio-frequency domain with equivalent spectral sweep times reaching microseconds. 

Despite many technological challenges, the importance of the MIR region has driven extensive research on OFC generation. In principle, many novel platforms offer octave-wide coverages, yet they still require bulky external pump lasers. This is in stark contrast to emerging battery-operated fully-integrated systems at telecom wavelengths~\cite{stern_battery-operated_2018}. An alternative way to generate MIR OFCs is exploiting the inherent gain nonlinearity (resonant in nature) in monolithic semiconductor lasers, which combine direct MIR OFC emission with high-power per comb line, wavelength tunability, and extremely small footprint~\cite{scalari_-chip_2019}. Moreover, semiconductor laser structures can amplify and absorb light at the same wavelength referred to as being bifunctional ~\cite{lotfiMonolithicallyIntegratedMidIR2016, schwarzMonolithicFrequencyComb2019,sterczewskiMidinfraredDualcombSpectroscopy2020}, which promises fully-integrated dual-comb sources and detectors defined photolitographically on the same chip. 

To date, MIR OFC generation in semiconductor lasers has been demonstrated in quantum cascade lasers (QCL)~\cite{hugiMidinfraredFrequencyComb2012}, type-II~\cite{bagheri_passively_2018, sterczewskiMultiheterodyneSpectroscopyUsing2017, schwarzMonolithicFrequencyComb2019}, and type-I~\cite{fengPassiveModeLocking252018} interband \emph{cascade} lasers (ICLs). Semiconductor laser combs offer terahertz-wide optical coverages and gigahertz repetition rates resulting in micro-to-milliwatt optical power per comb line. Unfortunately, the performance of the most widespread QCL comb platform drastically degrades at shorter wavelengths ($<$5 \textmu m), which relates to problematic dispersion compensation~\cite{luShortwaveQuantumCascade2018} and excitement of harmonic states with high modal sparsity~\cite{kazakovSelfstartingHarmonicFrequency2017}. Furthermore, the multi-watt power consumption hinders thermal and power management for QCL-based sensors. The recently demonstrated type-II ICL combs are an interesting alternative to QCLs with battery-compatible sub-watt power consumption and suitability for free-running DCS, yet their spectral properties in low-noise comb regimes remain plagued by the same issues as in short-wavelength QCLs; namely high modal sparsity due to harmonic/quasi-harmonic operation~\cite{sterczewskiMidinfraredDualcombSpectroscopy2019}. Mode grouping phenomena can be also added to this list~\cite{schwarzMonolithicFrequencyComb2019, sterczewskiMultiheterodyneSpectroscopyUsing2017}. Type-I ICL combs, in turn, have shown stable mode-locked comb generation only in external optical feedback configurations~\cite{feng2020passively} with reduced optical bandwidths and a loss of the monolithic design. Therefore, there is an exceedingly important niche for stable monolithic comb lasers to cover the organic C-H stretch band around 3~\textmu m, associated with the existence of life-related molecules.

Here, we address this demand and present a monolithic MIR OFC platform exploiting a GaSb-based type-I quantum well diode laser (QWDL) medium, whose comb operation was previously restricted only to the near-infrared~\cite{sterczewskiFrequencymodulatedDiodeLaser2020}. Our devices emitting around 3~\textmu m offer excellent comb coherence over a 1~THz optical bandwidth with almost flat single-lobed spectra. Free-running comb operation is enabled by the inherent gain nonlinearity, which yields a self frequency-modulated (FM) optical waveform with almost constant intensity and periodically swept instantaneous frequency. Sub-megahertz optical linewidths exhibited by the devices are fully compatible with free-running DCS. The diodes can be directly powered from AA batteries, which was used here to demonstrate a working, battery-operated MIR dual-comb source for mode-resolved spectroscopy. This platform also addresses the critical demand for spectrally-uniform stable comb sources without gaps in the emitted spectrum, mode clustering issues~\cite{sterczewski2021waveguiding} or operation in harmonic~\cite{piccardoHarmonicStateQuantum2018a}/quasi-harmonic states~\cite{bagheri_passively_2018,sterczewski_interband_2021}, which is a prerequisite for widespread adoption of chip-scale DCS beyond research laboratories. 

\section*{Results and Discussion}
\noindent \textbf{Device design:} The MIR QWDLs grown on a $n$-GaSb substrate employ type-I InGaAsSb/AlInGaAsSb quantum well active regions (see Experimental Section for structure and fabrication details). The QWDL is a typical single-section Fabry-P\'erot (FP) device, where the quasi-continuous-wave FM comb formation mechanism relies on the inherent gain nonlinearity~\cite{dong2020quantum, burghoffUnravelingOriginFrequency2020, opacakTheoryFrequencyModulated2019}. Therefore, on-chip saturable absorber sections obtained through ion implantation~\cite{bagheri_passively_2018, feng2020passively} are not needed at all. Self-starting, comb spectra develop simply under uniform DC bias of the entire single-section cavity in free-running conditions. Although it is expected that a multimode FP emitter may turn into a stable OFC, it is not guaranteed. The critical design consideration lies in ensuring sufficiently low group velocity dispersion (GVD) and suppressing modal leakage into the high-index substrate. These are motivated by our prior research on waveguiding and GVD properties of GaSb-based ICL devices~\cite{sterczewski2021waveguiding, vurgaftman2021toward}, which has identified the substrate and clad thicknesses as key contributors that significantly affect the spectral envelope shape and noise properties in a frequency comb mode. A weakly confined optical mode (i.e. with thin clad layers) may leak into the high-index GaSb substrate, reflect from the metallic contact, and return to the active region causing wavelength-dependent interference that spectrally modulates the gain and GVD profiles. This effect often triggered multi-spectral-lobe lasing that limited the optical bandwidth of GaSb-based OFCs and made continuous spectral tuning virtually impossible~\cite{sterczewski2021waveguiding, vurgaftman2021toward}. However, when under control, it may be used for vertical compensation of the intrinsically-high intracavity GVD, as opposed to longitudinal compensation via Gires-Turnois interferometers (GTI)~\cite{villares2016dispersion}. Here, by employing moderately thick clad layers (2~$\upmu$m) we induce weak ($\leq$0.1~cm$^{-1}$) quasi-periodic modulation of the broad gain spectrum, which causes an oscillatory phase shift due to the Kramers-Kronig relations. This in turn yields moderate oscillations in the gain-induced GVD profile that locally compensate the natively high positive material and waveguide GVD over sub-THz bandwidths. This allows low-dispersion regions to occur, which facilitate the formation of stable frequency combs via cascaded four-wave mixing (FWM) with broadband single-lobe spectra that can tune continuously. We would like to underline here, that in the view of recent theoretical works on FM comb formation discussing the role of nonzero GVD and Kerr nonlinearity on spectral bandwidth ~\cite{burghoffUnravelingOriginFrequency2020, opacakTheoryFrequencyModulated2019, beiser2021engineering}, we do not want the total GVD to be fully compensated. Instead, we want to suppress the highly positive waveguide and material GVDs, which do not permit stable comb formation in the QWDL devices when the total GVD is on the order of 1000's of fs$^2$/mm. We will start analyzing the comb properties first, while the tunability and dispersion characterization will be discussed later.

\begin{figure*}[t!]
	\centering
	\includegraphics[width=.9\linewidth]{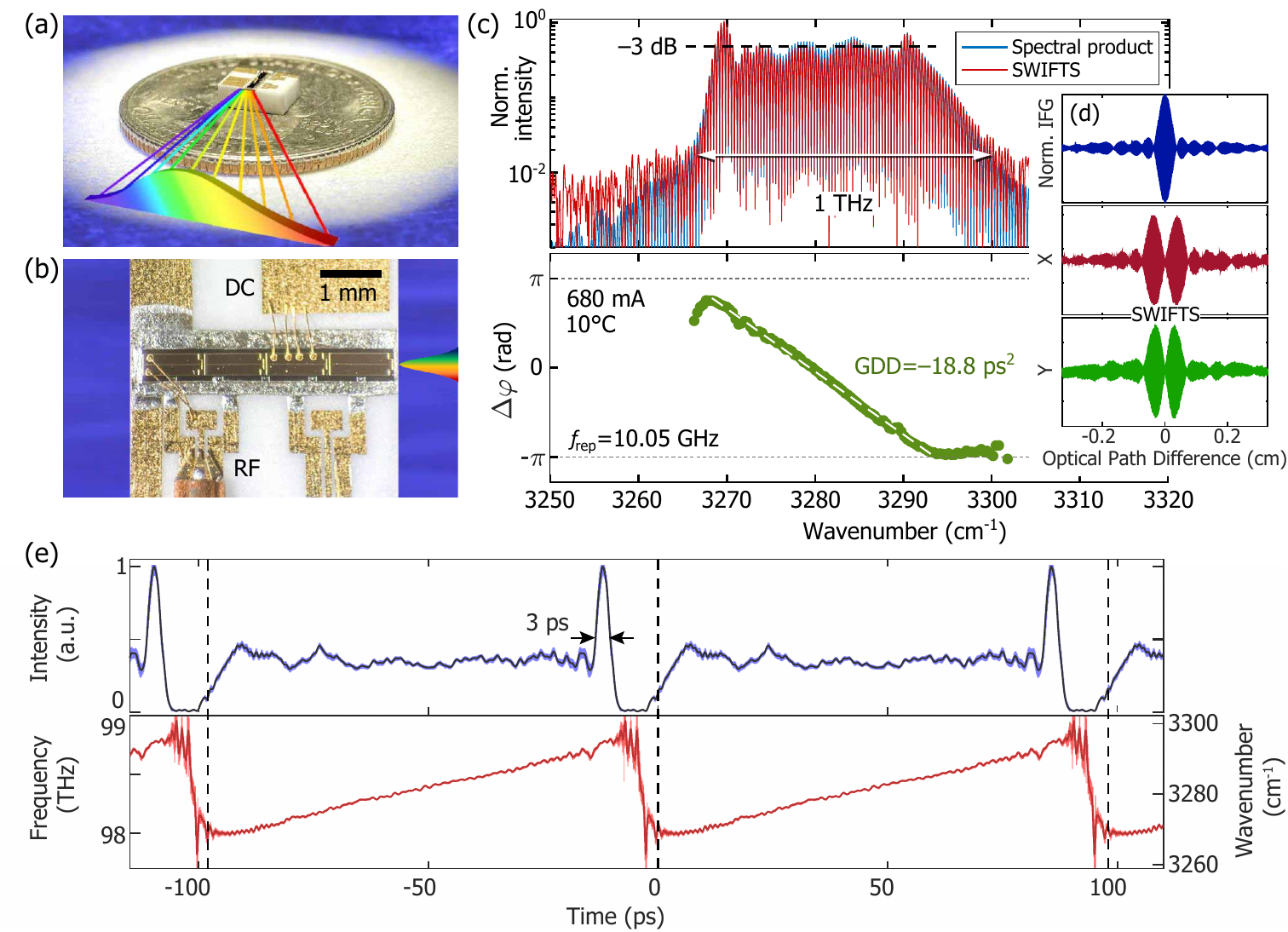}
	\caption{\textbf{Mid-infrared quantum well diode laser frequency combs.} \textbf{(a)} Picture of a device mounted onto a BeO submout. \textbf{(b)} Micrograph of the same device showing the radio-frequency (RF) signal extraction scheme with a microwave probe, and a DC bias pad. \textbf{(c)} Intensity and phase spectra obtained in the SWIFTS analysis at 680~mA of injection current. \textbf{(d)} Normal (DC) and quadrature (X/Y) SWIFTS IFGs proving frequency-modulated emission. \textbf{(e)} Reconstruction of the instantaneous intensity and frequency based on the modal intensities and phases. Shaded regions indicate a standard deviation obtained via a Monte–Carlo simulation, while dashed vertical lines split the reconstructed waveforms into periods.}
	\label{fig:fig1}
\end{figure*}

\vspace{0.2cm}
\noindent
\textbf{Comb properties:} To assess the frequency comb characteristics of MIR QWDLs, we mounted a 4~mm long single-section device  onto a radio-frequency (RF) compatible submount (\textbf{Figure~\ref{fig:fig1}a--b}), and performed a linear interferometric measurement known as shifted wave interference Fourier transform spectroscopy (SWIFTS)~\cite{burghoff_evaluating_2015}. This well-established technique allows for characterization of the coherence and relative phase between all comb lines in the spectrum (see Experimental Section) without resorting to nonlinear interferometry or intensity autocorrelation, which remain experimentally challenging at longer wavelengths. Instead, SWIFTS simultaneously records three interferograms (IFGs) from a fast linear photodetector: one normal (DC) and two microwave measured in quadrature at the repetition frequency -- $f_\mathrm{rep}$). 

The results of the analysis are plotted in \textbf{Figure~\ref{fig:fig1}c}. First, the device shows a remarkably uniform spectrum for a semiconductor OFC: the envelope has a single lobe with $\sim$3~dB intensity variations relative to the mean in the dominant part. This spectral pattern is observed at all currents exceeding the multimode operation threshold ($\sim$2.5$\times$ lasing threshold $J_\mathrm{th}$). 15 milliwats of optical power per facet are spread over 100 lines constituting the 20~dB optical bandwidth of 1~THz. Pronounced modal intensities around the spectral edges suggest strong frequency modulation, which is expected for single-section semiconductor laser devices that obtain comb emission through an interplay of spatial hole burning (SHB) and nonlinear FWM~\cite{opacakTheoryFrequencyModulated2019,burghoffUnravelingOriginFrequency2020}. The agreement between the spectral product (square root of the intensity spectrum product with an $f_\mathrm{rep}$-shifted replica), and the SWIFTS intensity spectrum proves that all lines contribute to comb emission. The intermodal phases splayed over almost a full range from $-\uppi$ to $\uppi$ indicate maximally chirped emission~\cite{opacakTheoryFrequencyModulated2019,burghoffUnravelingOriginFrequency2020} with an equivalent group delay dispersion of the electric field in the linear part of $-18.8$~ps$^2$. However, the shape deviates from a perfectly linearly chirped source, which manifests itself as oscillations and intermodal phase flattening close to the spectral edges. It is caused by two factors: the higher order dispersion of the generated waveform's electric field, and the coupling between the spectral amplitude and phase that converts minor intensity non-uniformities into phase oscillations. A minimum in the SWIFTS quadratures ($X/Y$) where the normal (DC) IFG has a maximum (\textbf{Figure~\ref{fig:fig1}d}) is a also signature of strong FM, like previously observed in near-infrared QWDL combs~\cite{sterczewskiFrequencymodulatedDiodeLaser2020}. 

Knowledge of the intermodal phases and modal intensities additionally enables us to reconstruct the temporal intensity and instantaneous frequency profiles, as shown in \textbf{Figure~\ref{fig:fig1}e}. A highly linear frequency sweep over almost 1~THz of optical bandwidth is accompanied by a nearly constant intensity except for 3-ps wide pulses occurring around the turnaround point. This amplitude pulsation seems to be a real feature confirmed by numerical simulations~\cite{opacakTheoryFrequencyModulated2019, burghoffUnravelingOriginFrequency2020} and other interferometric experiments~\cite{sterczewskiFrequencymodulatedDiodeLaser2020}.

\begin{figure*}[t!]
	\centering
	\includegraphics[width=0.95\linewidth]{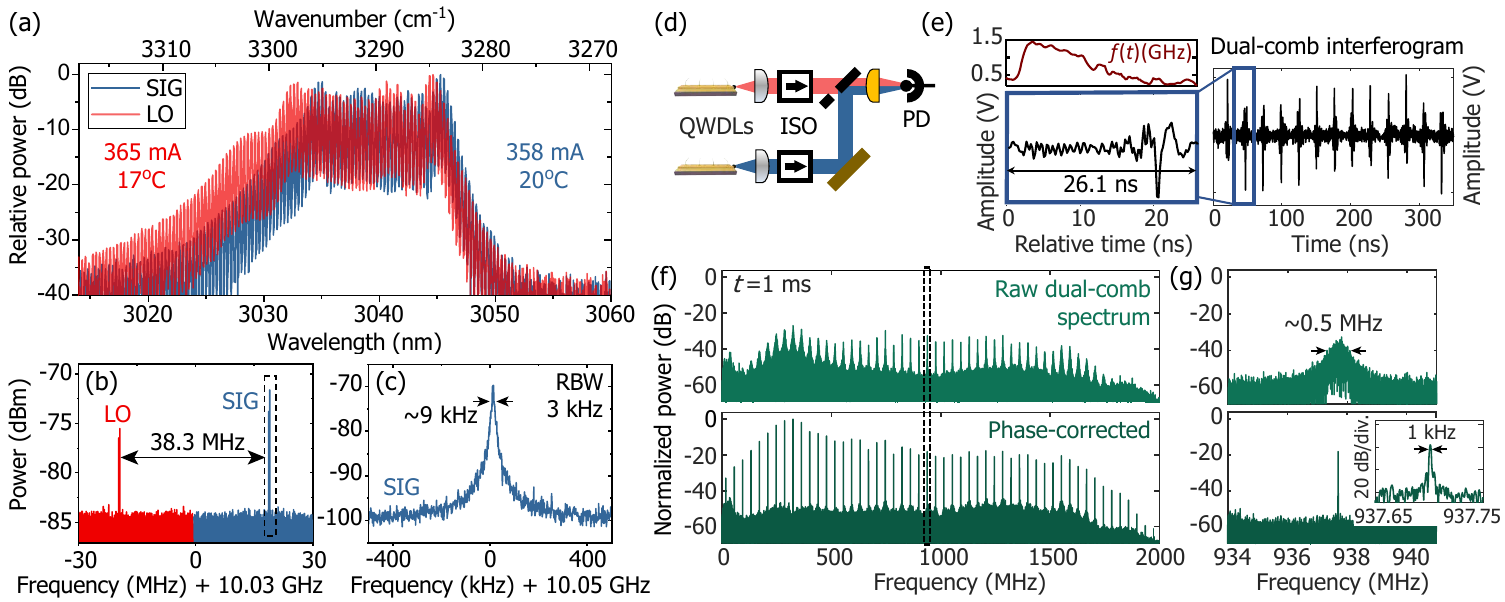}
	\caption{\textbf{Dual-comb beating of two free-running 3 \textmu m wavelength comb laser diodes.} \textbf{(a)} Optical spectra measured with a fiber-coupled spectrum analyzer. \textbf{(b)} Optically detected intermode beat notes. \textbf{(c)} Zoom on the signal (SIG) comb's intermode beatnote. \textbf{(d)} Experimental setup. Two QWDLs were optically-isolated (ISO) prior to combining on a photodetector (PD). \textbf{(e)} Dual-comb IFG measured from the photodetector's RF port along with a zoom on a single period. The retrieved instantaneous frequency $f(t)$ of the IFG shows strong linear chirp. \textbf{(f)} Raw (top) and digitally-corrected (bottom) dual-comb spectrum. The optical coverage is 520~GHz (limited by the oscilloscope). \textbf{(g)} Zoom on the center beat note before and after correction.}
	\label{fig:fig2}
\end{figure*}

\vspace{0.2cm}
\noindent \textbf{Suitability for free-running DCS:} Whereas SWIFTS gives direct access to relative phase coherence, it does not provide information about absolute comb stability i.e. optical linewidth. This is because accurate ($\sim$MHz) offset frequency retrieval $f_0$ would require an interferometer with a sub-kilometer optical path difference, which is impractical. Instead, to prove the excellent suitability of the MIR QWDL OFCs for demanding heterodyne measurements, we optically beat a pair of free-running, length-mismatched sources (several $\upmu$m) with a repetition rate mismatch of $\Delta f_\mathrm{rep}=38.3$~MHz. The light from the devices was spatially and spectrally overlapped onto a fast 3~\textmu m photodetector, and next the resulting DCS beating signal was recorded with an oscilloscope (see Experimental Section). Along with the lower-frequency DCS beating signal, we measured its high-frequency portion (near $f_\mathrm{rep}$) using an RF spectrum analyzer to quantify the timing stability. 

At injection currents of approximately $\sim$360~mA and bias voltages of $\sim$1.3~V, the devices developed 6~mW of optical power with $\sim$0.5~THz 10~dB bandwidths (\textbf{Figure~\ref{fig:fig2}a}) and kilohertz-wide intermode beat notes (Figure~\ref{fig:fig2}b,c). Such conditions were chosen to promote operation in the low GVD region (discussed later in the text) with narrow optical linewidths observable as sharp and intensive dual-comb lines. While at higher injection currents the devices should still possess the equidistance properties of an OFC as evident from the SWIFTS analysis, more phase noise was observed in the DCS signal seen as multi-megahertz-wide DCS beatnotes. To ensure the unperturbed operation of the combs, we used optical isolators for suppressing unwanted delayed optical feedback (Figure~\ref{fig:fig2}d).

Because the DCS technique gives access to the cross-correlation of the combs' electric fields, some of the temporal characteristics can be retrieved directly from the beating signal. To achieve that, we analyzed the time-domain electrical IFG shown in Figure~\ref{fig:fig2}e, which has a clear periodic structure with a non-zero offset frequency seen as an evolution of the waveform from frame to frame lasting 26.1~ns. Unexpectedly, it resembles a train of pulses followed by a noise-like oscillatory response. In fact, it is a highly chirped waveform that starts at a high frequency ($\sim$1.5~GHz, top panel), and gradually progresses toward lower frequencies. Due to the higher beat note amplitudes around 300--400~MHz, the IFG's amplitude increases as it approaches the lower frequency region, which potentially explains its pulsed-like shape. Nevertheless, phase flattening occurring close to the the high-wavenumber spectral edge (around 3295~cm$^{-1}$/3034~nm) may also contribute to the presence of weak pulsed features, as evident from the temporal intensity profile in Figure~\ref{fig:fig1}e.

The free-running dual-comb spectrum measured within 1~ms shows beat notes with signal-to-noise ratios (SNRs) reaching up to 25~dB (Figure~\ref{fig:fig2}f). Values higher by 10~dB were also possible, yet the detector and RF electronics responded in a nonlinear fashion producing intermodulation products. Interestingly, the beat note linewidth evolves across the spectrum. The central part near 1~GHz has $\sim$0.5~MHz-wide lines, which triples for lines located at the edges. Assuming uncorrelated operation of the devices, the corresponding optical linewidths lie in the 350--1000~kHz range. To better understand the nature of the line broadening phenomenon, we computationally extracted the instantaneous offset frequency difference $\Delta f_\mathrm{0}(t)$, and the repetition rate difference $\Delta f_\mathrm{rep}(t)$. Not only does it allow to prove that the entire DCS signal can be corrected by applying global phase correction using the two extracted signals~\cite{sterczewskiComputationalCoherentAveraging2019}, but also allows us to evaluate the degree of correlation between offset and timing fluctuations. We found that the two are anti-correlated with a correlation coefficient of $-0.4$. The symmetric broadening around the comb's center wavelength suggests the existence of a fixed point~\cite{walkerFrequencyDependenceFixed2007} there, where repetition rate (timing) fluctuations get partially compensated by offset frequency fluctuations. This explanation is in agreement with the recent studies on QCL combs~\cite{shehzadFrequencyNoiseCorrelation2020}. The obtainable spectral coverage in the DCS experiment reaches $\sim$520~GHz, partially limited by the oscilloscope electrical bandwidth, and the lower span of the signal (SIG) comb compared to the local oscillator (LO). 

\begin{figure}[t!]
	\centering
	\includegraphics[width=\columnwidth]{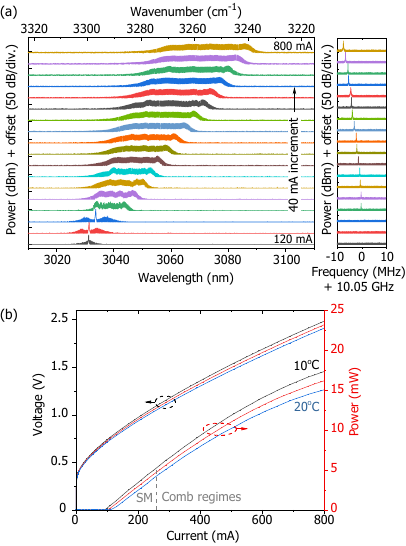}
	\caption{\textbf{Characteristics of the diode lasers.} \textbf{(a)} Injection current tuning of the optical and microwave spectrum at room temperature. \textbf{(b)} $L$-$I$-$V$ characteristics of the QWDL devices at different temperatures.}
	\label{fig:fig3}
\end{figure}

\vspace{0.2cm}
\noindent \textbf{Spectral properties and tunability:} For future application in spectroscopy, it is of practical relevance to characterize the spectral characteristics at different injection currents. This is because semiconductor microcombs often exhibit complex nonlinear dynamics that give rise to highly modulated optical spectra at some biases. Another motivation stems from the attractive perspective of going beyond the coarse GHz mode spacing to perform interleaved gap-less spectroscopy~\cite{gianella2020high}.

Unlike in many other comb platforms at MIR wavelengths, the spectral evolution exhibited by the QWDLs (\textbf{Figure~\ref{fig:fig3}}) is monotonic with remarkably smooth spectra, particularly at high higher ($>5J_\mathrm{th}$) injection currents. In contrast to QWDLs operating at much shorter wavelengths (2~$\upmu$m)~\cite{sterczewskiFrequencymodulatedDiodeLaser2020}, the MIR laser easily switches from single-mode to multi-mode (comb) operation at currents below 3$J_\mathrm{th}$ ($\sim$240~mA). This can be attributed to the longer emission wavelength, which facilitates spatial hole burning (SHB) governing the multi-longitudinal-mode emission mechanism. The so-called carrier grating formed inside the cavity periodically depletes the gain due to standing wave effects, with the period directly proportional to the modal wavelength. However, when the period is too long while carrier diffusion is too slow (i.e. at high injection currents when stimulated emission decreases the carrier lifetime), the laser cannot sustain single-mode lasing. Intuitively, at longer wavelengths this effect will trigger multi-mode lasing more easily, which is a prerequisite for an FP diode laser to become an OFC. 

To quantify this effect, we have employed a numerical SHB model proposed by Dong et al., which includes carrier saturation effects and a shortening of the carrier lifetime due to stimulated emission~\cite{dong2020quantum} (see Experimental Section for parameters used in the simulation). Although it is difficult to exactly predict the multimode emission threshold triggered by SHB, we can compare thresholds for devices employing the same QW medium but operating at different wavelengths~\cite{sterczewskiFrequencymodulatedDiodeLaser2020}. A critical condition is that the diffusion length ($L_D=\sqrt{D_\mathrm{a}\tau_\mathrm{cw}}$) be shorter than the quarter-wavelength inside the cavity ($\lambda/4n_0$)~\cite{vurgaftman2021toward}, where $D_\mathrm{a}$ is the ambipolar diffusion coefficient (10~cm$^2$/s based on Ref.~\cite{piskorski2016material}), $\tau_\mathrm{cw}$ is the carrier lifetime shortened via stimulated emission, and $n_0$ is the modal index. At a high intracavity optical power ($P_\mathrm{i}$ on the order of mW), the spontaneous carrier lifetime $\tau_\mathrm{sp}\approx3$~ns shortens to the picosecond range. In this regime, it may be approximated as $\tau_\mathrm{cw}\approx(2RP_\mathrm{i})^{-1}$, where $R$ is the recombination factor proportional to the wavelength. From a multimode operation standpoint, the relevant quantity is the ratio of the diffusion length to the quarter-wavelength $r_{L\lambda}=4n_0L_\mathrm{D}/\lambda$. Therefore, simply changing the operating wavelength by from $\lambda_0$ to $\lambda'$ with a similar heterostructure and device length (ignoring slight changes of other variables) should lower the longitudinal multimodeness power threshold $(\lambda_0/\lambda')^{-3/2}$ times. For example, a $3~\upmu$m FP diode laser should become multimode at intracavity powers $\sim1.8$ times lower than that operating at $2~\upmu$m. This prediction roughly agrees with experiments: 4~mm long $2.05~\upmu$m uncoated devices with a similar heterostucture developed multimode spectra at $\sim10$~mW of optical power per facet, while the $3.03~\upmu$m reported here required $\sim5$~mW. Including all relevant terms in the model (also accounting for carrier saturation, see Experimental Section for details), we find that the ratio of the diffusion length to the quarter wavelength $r_{L\lambda}$ in this medium that enables multimode emission (observed experimentally) must be lower than $\sim0.2$. In other words, the diffusion length $L_\mathrm{D}$ should be a small fraction of the quarter-wavelength rather simply equal to it. Once a multimode spectrum is developed, the inherent gain nonlinearity triggers FWM responsible for locking the modal phases, however, this process strongly depends on sufficiently low GVD~\cite{burghoffUnravelingOriginFrequency2020, vurgaftman2021toward, opacakTheoryFrequencyModulated2019}.

One of the key spectral characteristics of the presented MIR QWDL is the lack of spectral gaps or multi-lobed envelopes that often arise due to severe (uncontrolled) modal leakage into the substrate~\cite{bewley2004antimonide}. Although one cannot guarantee that MIR QWDLs display such characteristics in general, a dozen devices cleaved from the wafer (see Experimental Section for structure details) were tested and showed similar behavior. The center wavelength of the spectrum tunes in an almost linear fashion by $\sim$30~nm (32~cm$^{-1}$) over the entire dynamic range of comb operation ($I$=240--800~mA). In Fabry-P\'erot devices like the one discussed here, it is the temperature dependent red-shift of the maximum gain wavelength $\Delta \lambda_{g_\mathrm{max}}/\Delta T$ due to Joule heating (here equal to $\sim2$~nm/K) that predominantly governs the center wavelength $\lambda_\mathrm{c}$, while contributions from variations of the refractive and group indices are much smaller. The repetition rate (intermode) beat note does not exhibit any rapid jumps (Figure~\ref{fig:fig3}a, right) and also shows smooth tuning capabilities. In the comb operating range, it shifts by $\sim$8~MHz from the nominal value at highest injection currents. The most stable free-running combs are developed at currents ranging from 240 to 400~mA with intermode beat note linewidths below 10~kHz. Higher biases with broader linewidths, however, necessitate microwave injection locking~\cite{hillbrand2019coherent} to suppress the timing jitter, as in Figure~\ref{fig:fig1}. A full free-spectral-range (FSR) scan for spectral interleaving requires a change of injection current by $\sim$45~mA, which corresponds to an approximate optical tuning coefficient of $\Delta \nu/\Delta I=223$~MHz/mA.

\vspace{0.2cm}
\noindent
\textbf{Battery operation} From a power management perspective, the diode topology is highly beneficial~\cite{day2020simple,dong2020quantum}. The low compliance voltage below 2~V accompanied by moderate injection currents ($<$1~A) correspond to near-W electrical power consumption at room temperature when a dozen of mW of optical power per facet are produced. Highly appealing is that in the lowest-noise comb regime (240--400~mA), the compliance voltage is natively compatible with consumer-grade rechargeable NiMH batteries ($\sim$1.2~V). This feature encouraged us to implement a battery-operated mid-infrared free-running dual-comb spectrometer. Note that even under such biasing conditions, the amount of emitted optical power ($>$5~mW per comb per facet) perfectly fits real-life spectroscopic sensing requirements even with optical path enhancement techniques. 

\begin{figure}[t!]
	\centering
	\includegraphics[width=\columnwidth]{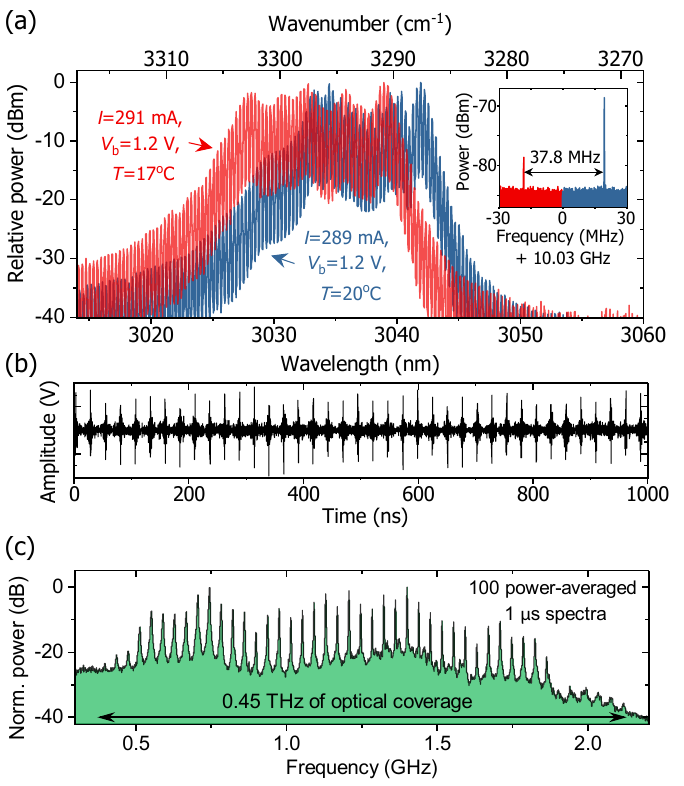}
	\caption{\textbf{Battery-operated mid-infrared dual-comb source.} \textbf{(a)} Optical spectra of the QWDLs biased at 1.2~V. \textbf{(b)} Dual-comb beating signal (IFG). \textbf{c}: Power-averaged dual-comb spectrum.}
	\label{fig:fig4}
\end{figure}

In the proof-of-principle experiment, each QWDL was biased directly from an AA battery without any control circuitry (also thermal) or microwave generators. Except for slightly lowering the spectral overlap and optical bandwidths due to the lower injection currents, other comb parameters remained virtually the same (\textbf{Figure~\ref{fig:fig4}a}) as in the current-stabilized experiment. The temporal structure of the dual-comb IFG preserves its regular spiking (Figure~\ref{fig:fig4}b). To lower the computational power requirements for a potential instrument, we did not perform any phase correction here prior to obtaining the frequency spectrum. Instead, the Welch periodogram averaging technique~\cite{welch1967use} was used. It relies on (incoherent) averaging of short-time power spectra to lower the uncertainty of the spectral amplitudes, yet without suppressing the noise floor. Such-processed spectrum (100 averages of 1~$\upmu$s-long signals) plotted in Figure~\ref{fig:fig4}c has a peak SNR of 20~dB, and a total optical coverage of 0.45~THz, which is highly attractive for embedded small-footprint DCS. 

Note that this proof-of-concept battery-operated experiment was performed in thermal non-equilibrium, i.e. the laser submount temperature was stabilized prior to starting the measurement to ensure mutual spectral overlap, and obviously drifted over time after the temperature control loop was turned off. From a portable instrument power budget standpoint, true steady-state operation implies that at least 50\% of the laser bias power should be added for cooling assuming a thermoelectric cooler is used for transferring the heat from the laser submount to a radiator. Fortunately, this level of consumed power does not violate the battery operation capability. 

\begin{figure}[t!]
	\centering
	\includegraphics[width=.9\columnwidth]{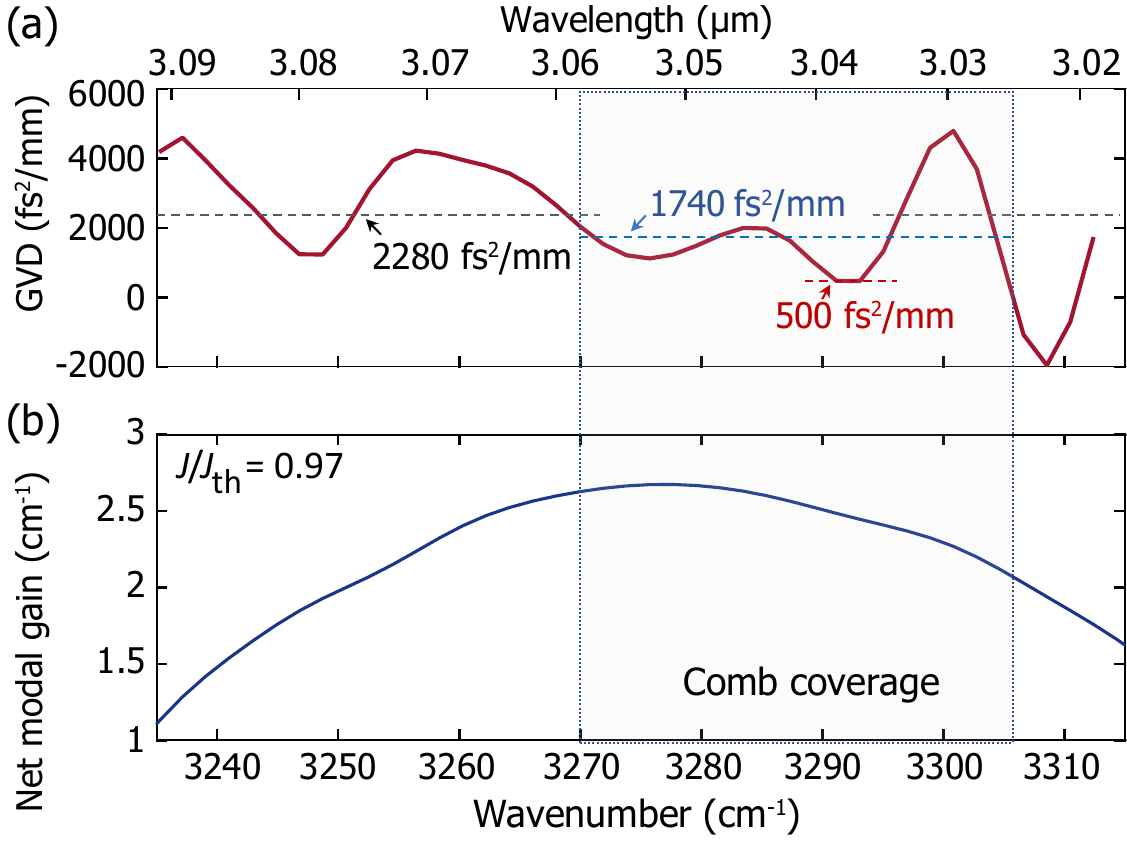}
	\caption{\textbf{GVD and gain measurement of the MIR QWDL comb (here SIG laser) at 0.97 of threshold current.} \textbf{(a)} GVD with the full-range mean of 2280~fs$^2$/mm, comb-range mean of 1740~fs$^2$/mm, and local minima of as low as 500~fs$^2$/mm. \textbf{(b)} Net modal gain.}
	\label{fig:fig5}
\end{figure}

\vspace{0.2cm}
\noindent \textbf{Gain and dispersion:} Another important aspect of the MIR QWDLs combs is their GVD, as it governs the ability to phase-lock the longitudinal modes via FWM~\cite{burghoffUnravelingOriginFrequency2020, vurgaftman2021toward, opacakTheoryFrequencyModulated2019}. We used the Fourier transform technique~\cite{hofstetter1999measurement} to study the gain and dispersion characteristics of the SIG device biased close to threshold (0.97$J_\mathrm{th}$) emitting light via amplified spontaneous emission (ASE). The measured GVD (\textbf{Figure~\ref{fig:fig5}a}) has a mean value of +2280~fs$^2$/mm in the 3.02--3.09~$\upmu$m range, which is comparable to other GaSb-based devices like ICLs~\cite{sterczewski2021waveguiding} in the 3~$\upmu$m region, yet still several times higher than long-wave ($>$7~$\upmu$m) InP-based QCL devices with negative material dispersion. One of the factors is the highly positive GVD of GaSb~\cite{adachi1989optical} ($>$1000~fs${^2}$/mm at 3.05~$\upmu$m), which gets further exacerbated in the device by contributions from the waveguide and gain. What is also observable in the GVD profile is its oscillatory quasi-periodic shape suggesting minor modal leakage into the high-index GaSb substrate. The period of 25--35~cm$^{-1}$ agrees with the prediction based on the substrate thickness (150$\pm20$~$\upmu$m)~\cite{bewley2004antimonide}. While it may be seen as a disadvantage, it allows certain spectral regions to exhibit lower dispersions than nominally without resonant contributions. Therefore, it acts as vertical dispersion compensation, which operates orthogonally to the more common GTI mirror obtained through multi-layer facet coating~\cite{villares2016dispersion}. For instance, in the comb operating range (3270--3305~cm$^{-1}$ / 3.025--3.058~$\upmu$m), the mean GVD decreases to 1740~fs$^2$/mm with local minima as low as 500~fs$^2$/mm. In fact, these low-dispersion regions appearing in the spectrum are not a measurement artifact -- they occur exactly where the most stable comb regimes with narrowest linewidths develop (around $\sim3292$~cm$^{-1}$). This justifies the need for future dispersion engineering~\cite{villares2016dispersion} to improve the bandwidth and stability of QWDL combs. The smooth net modal gain (Figure~\ref{fig:fig5}b) clearly supports broadband lasing and does not show any anomalies. Only weak ($\Delta g<0.1$~cm$^{-1}$) quasi-periodic sags appear possibly due to resonant (leaky) waveguide losses, yet without causing the laser to lase in multiple lobes. 

Incorporation of this vertical dispersion compensation scheme requires a careful (empirical) choice of clad thickness. Prior works on leaky mode GaSb ICLs have shown that an insufficiently thick clad layer yields a strongly modulated gain spectrum (as well as gain-induced GVD)~\cite{bewley2004antimonide, sterczewski2021waveguiding, sterczewski_interband_2021, vurgaftman2021toward}. Therefore, thicker clad layers are preferred for a less oscillatory profile. The same can in principle be obtained by making the bottom contact more diffuse. Another degree of freedom, which seems to be more controlled, is the gain modulation period $\Delta \tilde{\nu}$ [cm$^{-1}$] depending on the substrate thickness $d_\mathrm{sub}$, refractive index $n_\mathrm{GaSb}$, and modal index $n_\mathrm{m}$ as $\Delta \tilde{\nu}=(2\sqrt{n_\mathrm{GaSb}^2-n_\mathrm{m}^2} \, d_\mathrm{sub})^{-1}$ ~\cite{bewley2004antimonide,vurgaftman2021toward}.

\section*{Conclusion}
\noindent In conclusion, we have demonstrated a mid-infrared chip-scale frequency comb platform exploiting the type-I quantum well interband laser medium. The free-running stability exhibited by the lasers allows us to perform mode-resolved sub-THz-wide DCS around 3.04~$\upmu$m. The devices develop single-lobed smooth spectra with excellent tunability and show the potential for future gap-less interleaved spectroscopic measurements of spectrally-narrow molecular absorbers like acetylene (C$_2$H$_2$) or hydrogen cyanide (HCN). Particularly attractive is the low compliance voltage accompanied by low power consumption ($<1$~W), which was used to demonstrate the first battery-operated MIR dual-comb source. When merged with room-temperature interband cascade photodetectors~\cite{lotfiMonolithicallyIntegratedMidIR2016, schwarzMonolithicFrequencyComb2019, sterczewskiMidinfraredDualcombSpectroscopy2020}, the MIR QWDL devices promise dual-comb spectrometers compatible with field-deployable or even space instrument power budgets. The remaining difficulty lies in optimizing the signal processing side, which can potentially benefit from sample rate reduction techniques~\cite{sterczewski2020subsampling}. We also envision optical injection locking with a simple Vernier filter to synchronize the offset frequencies of the combs without an auxiliary single-mode laser, as recently proposed by Hillbrand et al.~\cite{hillbrandSynchronizationFrequencyCombs2022} to greatly simplify coherent averaging of DCS signals. Optimization of the microwave losses should enable us to perform strong ($>1$~W) microwave injection, which has shown to considerably broaden optical spectra in QCLs in addition to stabilizing the repetition rate~\cite{schneider2021controlling}. Although in this DCS demonstration we used two physical laser chips, two combs on the same chip are also possible. They would constitute a battery-operated MIR dual-comb generator with significant common-mode noise suppression~\cite{westberg2017mid} or even an on-chip dual-comb spectrometer~\cite{hillbrand2019coherent}. In addition to application in DCS, the QWDL's near-MHz optical linewidths are attractive for other, much simpler mode-resolved cavity-enhanced sensing techniques like Vernier spectroscopy~\cite{Sterczewski2021Vernier}. 
 
\section*{Methods}

\vspace{0.2cm}
\footnotesize
\setstretch{1.}
\noindent\textbf{3 \textmu m diodes}: The laser wafer was grown at National Research Council (NRC) Canada using solid-source molecular-beam epitaxy (MBE) on a $n$-GaSb (100) substrate. The 2-$\upmu$m-thick top and bottom cladding layers were formed by Te- and Be-doped Al$_{0.6}$Ga$_{0.4}$As$_{0.052}$Sb$_{0.948}$ layers. The laser active region consists of three 10.45-nm-thick In$_{0.55}$Ga$_{0.45}$As$_{0.21}$Sb$_{0.79}$ QWs separated by 30-nm-thick Al$_{0.2}$In$_{0.2}$Ga$_{0.6}$As$_{0.19}$Sb$_{0.81}$ quinary barriers. Heavily-doped GaSb layers are used for the contacts. 3~\textmu m wide ridges were defined photolitographically to ensure single spatial mode emission. The processed wafer was lapped down to a thickness of 150~$\upmu$m. No coating was deposited onto the facets -- they were left as cleaved. 

\vspace{0.2cm}

\noindent\textbf{SWIFTS}: The light emitted by the QWDL was collected using a black-diamond high-numerical-aperture anti-reflective (AR) coated lens, and next guided into a Bruker Vertex 80 FTIR spectrometer with an external photodetector. To prevent dynamic optical feedback that perturbed comb operation during an FTIR scan, an optical isolator was used.
The comb device with a natural repetition rate $f_\mathrm{rep}=10.05$~GHz was weakly (+6~dBm) injection locked using an external microwave generator, which additionally served a phase reference for a fast lock-in amplifier (Zurich Instruments UHFLI). The 10.05~GHz signal, however, was not used directly due to bandwidth limitations of the lock-in. An additional microwave signal offset by 20~MHz was synthesized to drive two mixers: one for the phase reference, and one for down-converting the detected optical signal. Together with the microwave signals, we recorded a DC (normal) IFG to compare the spectrum retrieved from microwave IFGs with the optical one. Note that the microwave injection signal was used only to serve as a reference for phase-sensitive detection in SWIFTS rather than to trigger comb generation, which starts by itself under DC bias of the cavity.  

\vspace{0.2cm}

\noindent\textbf{Dual-comb beating}: A pair of QWDL devices cleaved from the same wafer but different bars was slightly detuned in temperature by using thermoelectric coolers ($\Delta T=3$~K) to increase the repetition rate difference $\Delta f_\mathrm{rep}$ and maximize spectral overlap. In the first experiment, the injection currents and temperatures were controlled by a pair of D2-105 drivers (Vescent Photonics). In the battery-operated demonstration, however, no injection current control was used at all -- the diodes were supplied directly from the AA battery's terminals. The optical beams from the two combs were first guided through optical isolators, next attenuated using irises, and finally focused onto a fast thermoelectrically-cooled photodetector (PVI-4TE-3.4, VIGO) equipped with a 2.2~GHz-bandwidth transimpedance preamplifier. Excellent match between the detector responsivity characteristics and the source wavelength ensured high signal-to-noise ratio (SNR) of the electrical beating signal captured by a Lecroy WavePro7Zi-A oscilloscope configured in oversampling 11-bit resolution mode. Because of its limited 1.5~GHz electrical bandwidth, the microwave dual-comb spectra possess a rapid roll-off above 1500~MHz. 

\vspace{0.2cm}
\noindent\textbf{Spatial hole burning modeling}: The multimode operation threshold was estimated from the SHB model developed by Dong et al.~\cite{dong2020quantum}, which includes carrier saturation effects, and lifetime shortening due to stimulated emission. The quantity $\tau_\mathrm{cw}=(\eta I_\mathrm{in}/qN_\mathrm{qw}+1/\tau_\mathrm{sp}+2RP_\mathrm{i})^{-1}$ describes the equivalent lifetime, where $\tau_\mathrm{sp}$ is the spontaneous emission lifetime, $I_\mathrm{in}$ is the injection current, $\eta$ is the quantum efficiency, and $q$ is the elementary charge. $P_\mathrm{i}$ stands for the intracavity optical power approximated for uncoated devices as $P_\mathrm{i}\approx P_\mathrm{m}(1+r)/(1-r)$, where $r$ is the power reflection coefficient, and $P_\mathrm{m}$ is the measured power collected from a single facet. The recombination factor $R=2gL/\hbar \omega_0 N_\mathrm{qw}$ includes contributions from the modal gain coefficient $g$, cavity length $L$, angular transition frequency $\omega_0=2\mathrm{\pi}\nu_0$, and the effective number of states in the quantum well $N_\mathrm{qw}=D_\mathrm{r}^{2\mathrm{D}} \cdot \hbar \Gamma \cdot WLn_\mathrm{qw}h_\mathrm{qw}$. The first term of the latter is the reduced 2-D density of states $D_\mathrm{r}^{2\mathrm{D}}=m^*/(\mathrm{\pi} \hbar^2 h_\mathrm{qw})$, where $m^*$ is the electron effective mass, and $h_\mathrm{qw}$ is the QW height. The subsequent terms of $N_\mathrm{qw}$ are the homogeneous half-linewidth $\Gamma$ of the gain Lorentzian profile, the effective QW volume defined by the QW height times the number of QWs $n_\mathrm{qw}$, the waveguide width $W$, and the cavity length $L$. The values used in our simulations are as follows: $r=0.32$, $L=4$~mm, $g=50$~cm$^{-1}$, $m^*=0.045m_0$, $\Gamma=4$~meV$/\hbar$, $h_\mathrm{qw}=11$~nm, $W=3$~$\upmu$m, $n_0=3.6$, $\tau_\mathrm{sp}=3$~ns, $\eta=0.7$.

\vspace{0.2cm}
\noindent\textbf{GVD measurement}: The device labeled as SIG was biased below threshold at 30$^\circ$C. The collimated ASE light was guided into the step-scan Fourier Transform spectrometer in the same configuration as in the SWIFTS experiment. A low-bandwidth lock-in amplifier (SRS830, Stanford Instruments) was used to measure the interferometrically-modulated light (100~ms time constant). The GVD was retrieved from the second derivative of the roundtrip IFG burst phase spectrum. It should be noted that while the modal leakage effect shows high reproducibility of the modulation period (due to the same substrate thickness and modal indices), it displays lower in the location of peaks and troughs. Therefore, devices from the same wafer may possess low-GVD region slightly detuned from one another.

\footnotesize

\bibliographystyle{naturemag}
\bibliography{main}

\section*{Acknowledgements}
\footnotesize
\setstretch{1.}
\noindent
\noindent This work was supported under National Aeronautics and Space Agency’s (NASA) PICASSO program (106822 / 811073.02.24.01.85), and Research and Technology Development Spontaneous Concept Fund. It was in part performed at the Jet Propulsion Laboratory (JPL), California Institute of Technology, under contract with the NASA. L.~A. Sterczewski’s research was supported by an appointment to the NASA Postdoctoral Program at JPL, administered by Universities Space Research Association under contract with NASA. L.~A. Sterczewski acknowledges funding from the European Union's Horizon 2020 research and innovation programme under the Marie Skłodowska-Curie grant agreement No 101027721. Dr. Mark Dong at University of Michigan is acknowledged for fruitful discussions on spatial hole burning in diode lasers. We also acknowledge two anonymous reviewers for their helpful comments.

\section*{Author contributions}
\footnotesize
\setstretch{1.}
\noindent L.A.S., M.F., and M.B. conceived the idea. L.A.S. carried out the optical and electrical measurements, and analyzed the data. S.F. designed the laser structures. C.F. fabricated the devices reported here. L.A.S and M.B. wrote the manuscript with input from all authors. M.B. coordinated the project.

\section*{Conflict of interest}
\footnotesize
\setstretch{1.}
\noindent The authors declare no conflict of interest.

\section*{Data availability statement}
\footnotesize
\setstretch{1.}
\noindent Data are available in a public, open access repository. The data that support the findings of this study are openly available in figshare at \url{http://dx.doi.org/10.6084/m9.figshare.25057943}.
\end{document}